\pdfoutput=1

\documentclass[12pt]{article}
\usepackage{jheppub}

\usepackage{amsmath}
\usepackage{amssymb}
\usepackage{graphicx,color,slashed,braket}
\usepackage{bbm}
\usepackage{ifpdf}
\usepackage{comment}

\newcommand{\be}{\begin{equation}}
\newcommand{\ee}{\end{equation}}
\newcommand{\ba}{\begin{eqnarray}}
\newcommand{\ea}{\end{eqnarray}}
\newcommand{\nn}{\nonumber}
\newcommand{\lp}{\left(}
\newcommand{\rp}{\right)}

\renewcommand{\d}{\textrm{d}}
\newcommand{\e}{\textrm{e}}

\newcommand{\w}{\wedge}
\renewcommand{\a}{\alpha}
\renewcommand{\b}{\beta}
\newcommand{\N}{\mathcal{N}}

\def\rmi{{\rm i}}

\title{On the absence of supergravity solutions for localized, intersecting sources}

\author{Jacob Bardzell,}
\author{Kevin Federico,}
\author{Danielle Smith,}
\author{and Timm Wrase}

\affiliation{Department of Physics, Lehigh University, 16 Memorial Drive East, Bethlehem, PA 18018, USA\\}

\emailAdd{jdb319@lehigh.edu}
\emailAdd{kcf225@lehigh.edu}
\emailAdd{des219@lehigh.edu}
\emailAdd{timm.wrase@lehigh.edu}

\notoc

\abstract{
\noindent
For decades intersecting D-branes and O-planes have been playing a very important role in string phenomenology in the context of particle physics model building and in the context of flux compactifications. The corresponding supergravity equations are hard to solve so generically solutions only exist in a so-called smeared limit where the delta function sources are replaced by constants. We are showing here that supergravity solutions for two perpendicularly intersecting localized sources in flat space do not exist for a generic diagonal metric Ansatz. We show this for two intersecting sources with $p=1,2,3,4,5,6$ spatial dimensions that preserve 8 supercharges, and we allow for fully generic fluxes. 
}

\begin{document}

\maketitle

\newpage

\section{Introduction}\label{sec:introduction}

For more than two decades string theory compactifications with intersecting D-branes and O-planes have played an important role in string phenomenology. On the one hand, intersecting D-brane models are used to obtain particle physics models that can resemble the supersymmetric standard model and extension thereof, see for example the review article \cite{Blumenhagen:2005mu}. On the other hand, orientifold planes are needed in flux compactifications to partially break supersymmmetry and to provide a source of negative energy in the scalar potential, see for example \cite{Grana:2005jc, Blumenhagen:2006ci} for early review articles. For flux compactifications on toroidal orbifolds the orientifold planes generically intersect in the internal space. So, both settings lead to supergravity equations of motion that have localized sources that intersect in a non-trivial way.

For such intersecting sources one then has to solve the equations of motion for the electromagnetic field strengths that are being sourced. This is rather simple since the equations are linear and the field strengths for each individual source can simply be added up. However, Einstein's equations are non-linear and extremely hard to solve. This has led to the often-employed simplification of a so-called smearing of the sources over their transverse directions. Mathematically speaking one replaces the delta function sources with constants, which dramatically simplifies the equations of motion. If one does that, one would then have to try to understand how close such a smeared solution is to the actual localized solution one started with, which is not an easy question to answer \cite{Blaback:2010sj, Blaback:2011nz, Blaback:2011pn}.

One can of course try to solve the equations of motions for intersecting objects without smearing or by only partially smearing the sources. For example, one could smear only over the mutual transverse directions of all sources, or one smears the sources only over directions that are transverse to one and parallel to another, etc. This leads to a plethora of possibilities that are discussed for example in the review article \cite{Smith:2002wn} (see also \cite{Gauntlett:1997cv} for an earlier review article). The upshot of this endeavor is that fully localized solutions are known essentially only for parallel sources and in all other cases one has to do at least some partial smearing in order to solve the equations of motion. One exception is the case of two intersecting NS5-branes extending along $x^0, x^1, x^2, x^3, x^4, x^5$ and $x^0, x^1, x^6, x^7, x^8, x^9$ respectively (without any mutually transverse directions), see \cite{Gauntlett:1996pb} for a discussion of this solution.

Within the swampland program \cite{Vafa:2005ui} in string theory many flux compactifications have recently been revisited and scrutinized. In particular, flux compactifications of massive type IIA give rise to infinite families of weakly coupled 4d $\N=1$ AdS vacua \cite{DeWolfe:2005uu, Camara:2005dc}. The viability of these solution was questioned for example by the AdS swampland conjecture \cite{Lust:2019zwm}. One criticism pertaining to these type IIA flux compactifications is that they are using smeared orientifolds planes, i.e., the full 10d supergravity equations of motion have not been explicitly solved \cite{Acharya:2006ne}. Two papers recently revisited this problem \cite{Junghans:2020acz, Marchesano:2020qvg} and found approximate solution with localized sources (see also \cite{Cribiori:2021djm, Collins:2022nux, Apers:2022tfm, Emelin:2022cac, VanHemelryck:2022ynr, Apers:2022vfp, Shiu:2022oti, Cribiori:2023ihv, Farakos:2023nms, Junghans:2023lpo, Carrasco:2023hta, Tringas:2023vzn, Andriot:2023fss, Junghans:2023yue, Farakos:2023wps} for closely related recent work). These approximate solutions in \cite{Junghans:2020acz, Marchesano:2020qvg} arose from an expansion in the large $F_4$-flux quanta and they capture the leading order backreaction of the localized orientifold planes. However, at this order the actual effects of the intersection of the O-planes is not taken into account. It would therefore be extremely important to extend these approximate solutions to higher order. However, given the importance of intersecting sources in many parts of string theory, a broader approach is also certainly warranted.

In this paper we study the equations of motion for two localized D$p$-branes or O$p$-planes in flat space. We take them to intersect perpendicular with four Neumann/Dirichlet directions and $(p-2)$ common directions (often denoted D$p\perp\,$D$p\,(p-2)$). This means the setup preserves 8 supercharges, which allows us study the SUSY transformations of the fermions. Demanding that these vanish, as required for a supersymmetric solution, we find that a fully localized solution cannot exist for a generic diagonal metric Ansatz, even when allowing for fully generic fluxes. While this might come as a surprise, similar results were previously obtained. For example, it was shown in \cite{Patino:2004cz} that no solution can exist for localized, intersecting D3/D5-branes.

The outline of the paper is as follows: In section \ref{sec:review} we review the supergravity solution for a single source. Then we discuss two perpendicularly intersecting objects in section \ref{sec:2sources} and show that the corresponding equations of motion have no solution. In section \ref{sec:Outlook} we discuss our findings and provide an outlook on important open questions.

\section{Review of a single source}\label{sec:review}
In this section we will solve the equations of motion of type II supergravity coupled to a stack of D$p$-branes or an O$p$-plane in 10d flat space. Such a solution is textbook material \cite{Johnson:2003glb} but we review it here to set up our notation and to remind the reader of some features that will be important in the next section.

\subsection{Type II supergravity}\label{ssec:EOMs}
We are using the notation and conventions of \cite{Blaback:2010sj} but we will change to string frame. The trace reversed Einstein equations are given by
\ba\label{eq:GR}
R_{ab} &=& - 2\nabla_a \partial_b\phi+ \frac14 g_{ab} \lp 2 g^{cd}\partial_c \phi \partial_d \phi -\nabla^2 \phi \rp +\frac12 |H|^2_{ab} -\frac18 g_{ab} |H|^2\\
&&+\sum_{n\leq 5} \e^{2\phi}\lp \tfrac{1}{2(1+\delta_{n5})} |F_n|^2_{ab} - \tfrac{n-1}{16(1+\delta_{n5})} g_{ab} |F_n|^2 \rp +\frac12 \e^\phi\lp T_{ab}^{loc} -\frac18 g_{ab} T^{loc} \rp \,.\nn
\ea
The sum over $n$ includes all even/odd numbers from 0 to 5 for IIA/IIB. The $\delta_{n5}$ is the Kronecker delta, and squares of $q$-forms are defined via $|A|^2_{\a\b} = \tfrac{1}{(q-1)!}A_{\a a_2\ldots a_q} {A_\b}^{a_2\ldots a_q}$, $|A|^2 = \tfrac{1}{q!}A_{a_1\ldots a_q} A^{a_1\ldots a_q}$. We restrict to parallel (stacks) of D$p$-branes or O$p$-planes so that the local stress tensor is given by
\be
T_{\mu\nu}^{loc} = \mu_p \, g_{\mu\nu}\, \delta(p)\,.
\ee
Here $\mu_p$ is negative for D$p$-branes and positive for an O$p$-plane.\footnote{While we do not need the exact values, the charge and tension of a stack of $N_p$ D$p$-branes is $-N_p \tilde{\mu}_p = -N_p (2\pi \sqrt{\alpha'})^{-p}/\sqrt{\alpha'}$. The charge and tension of an O$p$-plane is $-2^{p-5}\tilde{\mu}_p$ in the quotient space. The quantity appearing in our equations is $\mu_p = -N_p 2\kappa_{10}^2 \tilde{\mu}_p =- N_p (2\pi\sqrt{\alpha'})^{7-p}$ for a stack of D$p$-branes and $\mu_p = 2^{p-5}(2\pi\sqrt{\alpha'})^{7-p}$ for an O$p$-plane.} $\delta(p)$ denotes a delta function that localizes us on the $p+1$ dimensional world volume of the source. For multiple parallel D$p$-branes or O$p$-planes $\delta(p)$ should be understood as a sum of $\delta$-functions. $\mu,\nu$ are denoting the directions along the worldvolume of the source and $g_{\mu\nu}$ is the pullback of the spacetime metric $g_{ab}$ to the worldvolume of the source.

The equation of motion for the dilaton is given by
\be
\nabla^2 \phi =  2g^{ab}\partial_a\phi \partial_b \phi -\frac{1}{2} |H|^2 +\sum_{n < 5} \tfrac{5-n}{4} \e^{2\phi} |F_n|^2 - \tfrac{p-3}{4} \e^{\phi} \mu_p \delta(p)\,. 
\ee
Plugging the above into equation \eqref{eq:GR}, we find that it simplifies to
\ba
R_{ab} &=& - 2 \nabla_a \partial_b \phi + \frac12 |H|^2_{ab} + \sum_{n \leq 5}  \e^{2 \phi} \lp   \tfrac{1}{2(1+ \delta_{n5})} |F_n|^2_{ab} - \tfrac{1}{4(1 + \delta_{n5})} g_{ab} |F_n|^2 \rp \cr
&&+ \frac12 \e^\phi\lp T_{ab}^{loc} -\frac12 g_{ab} \mu_p \delta(p) \rp\,.
\ea
In the absence of NS5-branes, the Bianchi identities for the field strengths are
\ba\label{eq:Bianchi}
d H &=& 0\,,\cr 
d F_n &=& H\w F_{n-2} - \mu_{8-n} \delta_{n+1}(8-n)\,,
\ea
where $\delta_{n+1}(8-n)$ is a shorthand notation for the delta function $\delta(8-n)$ multiplied by a normalized $(n+1)$ volume form transverse to the source. 

The equations of motion for the gauge fields in the absence of NSNS sources are given by
\ba\label{eq:EM}
\d\lp \e^{-2\phi} \star H \rp &=& -\frac12 \sum_{n\leq10} \star F_n \w F_{n-2}\,,\cr
d \lp \star F_n \rp &=& - H \w \star F_{n+2} - (-1)^{\frac{n(n-1)}{2}} \mu_{n-2} \delta_{11-n}(n-2)\,.
\ea
The equations of motion for the RR fields can be obtained from the Bianchi identities in equation \eqref{eq:Bianchi} by using that $F_n=(-1)^{\frac{(n-1)(n-2)}{2}} \star F_{10-n}$.

For supersymmetric solutions one has to require that the SUSY transformations of the fermions vanish. This provides a simpler set of first order equations that often completely fixes the system and thereby automatically solves the Einstein and dilaton equations. We use the conventions of \cite{Bergshoeff:2001pv, Grana:2005jc} so that the transformations of the gravitino and gaugino are given by
\begin{align}\label{eq:spinortrafos}
\delta_\epsilon \psi_a &= \lp \partial_a +\frac14 \underline{\omega_a} + \frac14 \underline{H_a} \mathcal{P}\rp \epsilon+ \frac{1}{8} e^\phi \sum_n \frac{1}{1 + \delta_{n5}} \,\underline{F_n} \Gamma_a \mathcal{P}_n \epsilon\,,\cr
\delta_\epsilon \lambda &= \left( \underline{\partial \phi} + \frac14 \underline{H} \mathcal{P}\rp \epsilon + \frac14 e^\phi \sum_n (-1)^n (5-n) \underline{F_n} \mathcal{P}_n \epsilon\,.
\end{align}
The sum over $n$ includes all even/odd numbers from 0 to 5 for IIA/IIB. As above $a=0,1,\ldots, 9$ is a curved space index and we denote the corresponding tangent space indices as $A,B=0,1,\ldots, 9$. The underlined quantities are given by
\ba
\underline{\omega_a} &= {\omega_a}^{AB} \Gamma_{AB}\,, \quad &\underline{H_a} =\frac12 H_{abc} \Gamma^{bc}\,, \quad \underline{H} =\frac{1}{3!} H_{abc} \Gamma^{abc}\,, \cr
\underline{F_n}&=\frac{1}{n!} F_{a_1\ldots a_n} \Gamma^{a_1\ldots a_n}\,, \quad &\underline{\partial \phi}=\partial_a \phi \Gamma^a\,,
\ea
where $\Gamma^{a_1 a_2\ldots a_n} = \Gamma^{[a_1} \Gamma^{a_2} \ldots \Gamma^{a_n]}$, and we also define $\Gamma_{10} = \Gamma_{012\ldots 9}$. Furthermore, we have that
\ba
&\mathcal{P} = \Gamma_{10} \quad \text{in IIA,} \qquad &\mathcal{P}=-\sigma_3 \quad \text{in IIB,} \\
&\mathcal{P}_n = (\Gamma_{10})^{\frac{n}{2}} \quad \text{in IIA,} \qquad &\mathcal{P}_n=\sigma_1 \text{ for } \frac{n+1}{2} \text{ even, }\quad \rmi \sigma_2 \text{ for } \frac{n+1}{2} \text{ odd,} \quad \text{in IIB}\,.\qquad \nn
\ea
The spinor $\epsilon$ in type IIA has 32 real components, which could be split into two 16 component Majorana-Weyl spinors with opposite chiralities: $\Gamma_{10} \epsilon_1 = \epsilon_1$, $\Gamma_{10} \epsilon_2 =-\epsilon_2$. For IIB $\epsilon = (\epsilon_1, \epsilon_2)^T$ is a doublet of two 16 component Majorana-Weyl spinors with positive chirality so that $\Gamma_{10} \epsilon_i = \epsilon_i$. The Pauli matrices $\sigma_i$ above act on this doublet.

In the presence of D$p$-branes along the first $p+1$ directions or when doing the corresponding orientifold projection we break half of the supersymmetry via the following projection (involving the flat space $\Gamma$-matrices)
\be
\epsilon_2 = \Gamma_{01\ldots p} \epsilon_1\,.
\ee

\subsection{A single $p$-dimensional source}\label{ssec:singlesource}
We consider first a single O$p$-plane or a stack of D$p$-branes. These localized objects are magnetic sources for $F_{8-p}$ due to their Chern-Simons coupling to $C_{p+1}$. So, the only sourced RR-field is $F_{8-p} = \star F_{p+2}$. We can set all other RR-fields and the NSNS-flux $H$ equal to zero.

We can choose our coordinates in such a way that the O$p$-plane or the stack of D$p$-branes extend along $x^\mu$ with $\mu=0,1,\ldots, p$ and are located at the origin in the transverse directions $x^i=0$, for $i=p+1,p+2,\ldots,9$. This then preserves an $SO(p,1)\times SO(9-p)$ symmetry group, where the first $SO(p,1)$ factor is enhanced to the full Poincar\'e group. The most general metric Ansatz that is compatible with these symmetries is
\begin{equation}
 g= e^{2A_1(r)} \eta_{\mu\nu} dx^\mu dx^\nu + e^{2A_2(r)} \delta_{ij} dx^i dx^j\,. 
\end{equation}
Here $e^{2A_1(r)}$ and $e^{2A_2(r)}$ can only depend on $r=\sqrt{(x^{p+1})^2+\ldots+(x^9)^2}$, the overall transverse distance from the localized source.

The solution to the equations above in subsection \ref{ssec:EOMs} can be found in the textbook \cite[(10.38)]{Johnson:2003glb} and we write it as
\begin{align}
    e^{-4A_1(r)} &= e^{4A_2(r)} = 1 -  \frac{\tilde{\mu}_p}{r^{7-p}} \,,\label{eq:eA}\\
    e^{\phi(r)} &= e^{\phi_0 +(p-3) A_1(r)} \,,\label{eq:ephi}\\
    C_{p+1}(r) &= \lp 1-e^{4A_1(r)} \rp e^{-\phi_0} \ dx^0 \w dx^1\w \ldots \w dx^p \,.
\end{align}
Here $e^{\phi_0}$ is the asymptotic value of the dilaton infinitely far away from the source. We fixed the metric to be asymptotically Minkowski and we chose $C_{p+1}(r)$ to asymptotically vanish. We also defined $\tilde{\mu}_p = (-1)^{p+1} e^{ \phi_0}\frac{\Gamma(\frac{9-p}{2})}{2(7-p)\pi^{\frac{9-p}{2}}}\mu_p$.

Note that due to the minus sign in equation \eqref{eq:eA} and the fact that $\tilde{\mu}_p$ is positive for an O$p$-plane, there is actually a singularity at a finite distance $r={\tilde{\mu}_p}^{\frac{1}{7-p}}$ from the O$p$-plane. This singularity is at a distance that is of the order of the string length, $l_s = 2\pi \sqrt{\alpha'}$. At this point stringy corrections modify the equations of motion and remove this singularity.

Since we will need this later, we derive here explicitly the solution to the non-trivial Bianchi identity (cf. equation \eqref{eq:Bianchi}). We rewrite it using the transverse metric determinant $g_{9-p} = e^{2(9-p)A_2}$ as follows
\begin{align}
  d F_{8-p} &= - \mu_{p}\, \delta_{9-p}(p)\cr  
  &= - \mu_{p}\,\delta_{9-p}(p) \star_{9-p} 1\cr 
  &= - \mu_{p}\, \frac{1}{\sqrt{g_{9-p}}} \,\delta(x^{p+1})\delta(x^{p+2})...\delta(x^9) \,\sqrt{g_{9-p}} \ dx^{p+1} \wedge dx^{p+2}\wedge... \wedge dx^9\cr 
   &= - \mu_{p}\, \delta(x^{p+1})\delta(x^{p+2})...\delta(x^9)\,dx^{p+1} \wedge dx^{p+2}\wedge... \wedge dx^9\cr 
   &= - \mu_{p} \, \tilde{\delta}(\vec{r}) \, \tilde{\star}_{9-p} 1\,.
\end{align}
The tilde indicates that we are working with the flat space metric so there is no warp factor dependence anymore. The solution is given by
\begin{equation}\label{eq:SolutionBianchiSingleSource}
    F_{8-p} =  \tilde{\star}_{9-p}\, d \left(\frac{\tilde{\mu}_{p}}{r^{7-p}}\right) \,,
\end{equation}
since
\begin{align}
    d F_{8-p}&= \d\tilde{\star}_{9-p} d \left(\frac{\tilde{\mu}_{p}}{r^{7-p}}\right) \cr
    &= (-1)^p (\tilde{\star}_{9-p} 1 )\tilde{\nabla}^2 \left(\frac{\tilde{\mu}_{p}}{r^{7-p}}\right) \cr
    &= (-1)^{p+1} (\tilde{\star}_{9-p} 1 ) \tilde{\mu}_{p}\frac{2(7-p)\pi^{\frac{9-p}{2}}}{\Gamma(\frac{9-p}{2})} \tilde{\delta}(\vec{r}) \cr
     &= (\tilde{\star}_{9-p} 1 )\, \mu_{p}\,\tilde{\delta}(\vec{r})\,.
\end{align}

Summarizing, we see that it is possible to solve the supergravity equations exactly for a single source. Similarly, one can solve the equations of motion for parallel sources that are located not necessarily at $\vec{r}=0$ but at different positions $\vec{r}_0^{\, (\alpha)}$, $\alpha=1,2,\ldots$. In this case we can simply add up the individual solutions for each source and the solution is given by
\begin{align}\label{eq:parallelsources}
    e^{-4A_1(\vec{r})} &= e^{4A_2(\vec{r})} = 1 - \sum_{\alpha} \frac{\tilde{\mu}_p^{(\alpha)}}{\big|\vec{r}-\vec{r}_0^{\, (\alpha)}\big|^{7-p}} \,,\cr
    e^{\phi(\vec{r})} &= e^{\phi_0 +(p-3) A_1(\vec{r})} \,,\cr
    C_{p+1}(\vec{r}) &= \lp 1-e^{4A_1(\vec{r})} \rp e^{-\phi_0} \ dx^0 \w dx^1\w \ldots \w dx^p \,.
\end{align}
Note that the Bianchi identities in equation \eqref{eq:Bianchi} are linear and we can always simply add up the field strengths for any arbitrarily complicated configuration of sources. However, it is highly unusual, and special to this case of parallel sources, that the non-linear general relativity equation in \eqref{eq:GR} is also solved if we simply add up solutions.

\section{Two perpendicularly intersecting sources}\label{sec:2sources}

In this section we want to solve the equations of motion for two perpendicularly intersecting $p$-dimensional sources in flat space. These could be either two O$p$-planes or two stacks of D$p$-branes or one of each. We restrict to $1 \le p \le 6 $ so that we can have four directions that are along one of the objects and transverse to the other and there is at least one common transverse direction. The configuration that preserves eight supercharges in flat space is shown below.

\begin{center}
\begin{tabular}{|l|c|c|c|c|c|c|c|c|c|c|c|c|}\hline
     Spacetime directions & 0 & ... & $p-2$ & $p-1$ & $p$ & $p+1$ & $p+2$& $p+3$ & ... & 9 \\\hline
     First source & $\times$ & $\times$& $\times$ & $\times$ & $\times$ & -& - & - & - & -  \\\hline
     Second source & $\times$ & $\times$ & $\times$ & - & - & $\times$ & $\times$ & -& - & - \\\hline
\end{tabular}
\end{center}

The above intersecting sources respect an $SO(p-2,1)\times SO(2) \times SO(2) \times SO(7-p)$ symmetry.\footnote{For the special case of $p=1$ there are no directions common to both sources and therefore no $SO(p-2,1)$ factor. However, this does not affect our reasoning.} The first $SO(p-2,1)$ group is actually enhanced to the full Poincar\'e group. This symmetry group together with the specific source configuration shown above allows the metric (warp factors) to only depend on $\rho_1=\sqrt{(x^{p+1})^2+(x^{p+2})^2}$, $\rho_2=\sqrt{(x^{p-1})^2+(x^p)^2}$ and $\rho_T=\sqrt{(x^{p+3})^2 + ... + (x^9)^2}$. We make the following diagonal metric Ansatz
\begin{align}\label{eq:metric}
    ds^2 &= e^{2A_1(\rho_1,\rho_2,\rho_T)} \eta_{\mu\nu} dx^\mu dx^\nu + e^{2A_2(\rho_1,\rho_2,\rho_T)} \left((dx^{p-1})^2+(dx^p)^2\right)\\
    &+ e^{2A_3(\rho_1,\rho_2,\rho_T)} \left((dx^{p+1})^2+(dx^{p+2})^2\right)+ e^{2A_4(\rho_1,\rho_2,\rho_T)} \lp (dx^{p+3})^2 + ... + (dx^{9})^2 \rp\,,\nonumber
\end{align}
with $\mu,\nu=0,1,\ldots, p-2$. Poincar\'e invariance ensures that the first part of the metric is generic and there cannot be any off-diagonal terms like for example $g_{\mu \rho_1} dx^\mu d\rho_1$ since there are no invariant constant vectors of $SO(p-2,1)$. Non-constant vectors like $\eta_{\mu\nu}x^\mu dx^\nu$ are forbidden by translational invariance. However, in general there could be terms involving $d\rho_1d\rho_T$, etc. and also terms involving the corresponding angles $d\theta_1$ and $d\theta_2$ when going to polar coordinates, $(x^{p+1},x^{p+2}) \to (\rho_1, \theta_1)$ and $(x^{p-1},x^p) \to (\rho_2, \theta_2)$. Here we are restricting to a diagonal metric to make the problem tractable. Since the source setup is invariant under the exchanges $x^{p-1} \leftrightarrow x^p$ and $x^{p+1} \leftrightarrow x^{p+2}$ we can impose the same symmetry on the metric Ansatz, making equation \eqref{eq:metric} the most general diagonal metric Ansatz compatible with the source configuration.

We choose to work with Cartesian coordinates that have the following property that will be important below
\begin{align} \label{eq:derivatives}
    \partial_{x^{p-1}} e^{2A_n(\rho_1,\rho_2,\rho_T)} &= \frac{x^{p-1}}{\rho_2} \partial_{\rho_2} e^{2A_n(\rho_1,\rho_2,\rho_T)}\,,\cr
    \partial_{x^p} e^{2A_n(\rho_1,\rho_2,\rho_T)} &= \frac{x^p}{\rho_2} \partial_{\rho_2} e^{2A_n(\rho_1,\rho_2,\rho_T)}\,,\cr
    \partial_{x^{p+1}} e^{2A_n(\rho_1,\rho_2,\rho_T)} &= \frac{x^{p+1}}{\rho_1} \partial_{\rho_1} e^{2A_n(\rho_1,\rho_2,\rho_T)}\,,\cr
    \partial_{x^{p+2}} e^{2A_n(\rho_1,\rho_2,\rho_T)} &= \frac{x^{p+2}}{\rho_1} \partial_{\rho_1} e^{2A_n(\rho_1,\rho_2,\rho_T)}\,.
\end{align}
For the dilaton the most general Ansatz is $\phi = \phi(\rho_1,\rho_2,\rho_T)$. We also define the transverse coordinates for the two O-planes
\begin{align}
    r_1 &= \sqrt{\rho_1^2+ \rho_T^2} = \sqrt{(x^{p+1})^2+(x^{p+2})^2+(x^{p+3})^2+...+(x^9)^2}\,,\cr
    r_2 &= \sqrt{\rho_2^2+\rho_T^2} = \sqrt{(x^{p-1})^2+(x^p)^2+(x^{p+3})^2+...+(x^9)^2}\,.
\end{align}

Using the metric Ansatz as given in equation \eqref{eq:metric}, we seek the solution for the above source configuration. We first solve the linear Bianchi identity (cf. equation \eqref{eq:Bianchi})
\begin{equation}\label{eq:2sourcesBianchi}
        d F_{8-p} = - \mu_{p}^{(1)} \delta_{9-p}^{(1)}(p_1) - \mu_{p}^{(2)} \delta_{9-p}^{(2)}(p_2)\,.
\end{equation}
We solve the above equation by writing $F_{8-p}=F_{8-p}^{(1)} + F_{8-p}^{(2)}+F_{8-p}^{(c)}$, where $dF_{8-p}^{(c)}=0$ is closed\footnote{We are indebted to Daniel Junghans for pointing out this additional closed piece in $F_{8-p}$.} and
\begin{equation}
    d F_{8-p}^{(1)} = - \mu_{p}^{(1)} \delta_{9-p}^{(1)}(p_1) \, \qquad \text{and} \qquad d F_{8-p}^{(2)} = - \mu_{p}^{(2)} \delta_{9-p}^{(2)}(p_2) \,.
\end{equation}
So, this manifests the linearity of the electromagnetic equations and allows us to simply add up the two fields strengths for the two sources, i.e., we can add up the results for two single sources in flat space. The solution for the first source is (cf. equation \eqref{eq:SolutionBianchiSingleSource})
\begin{equation}\label{eq:F21}
    F_{8-p}^{(1)} =  \tilde{\star}_{9-p}^{(1)} d \left(\frac{\tilde{\mu}_{p}^{(1)}}{r_1^{7-p}}\right).
\end{equation}
From equation \eqref{eq:F21} we can read off the non-zero components of $F_{8-p}^{(1)}$
\begin{align}\label{eq:F21expl}
    F_{8-p}^{(1)} =& \tilde{\star}_{9-p}^{(1)} d \left(\frac{\tilde{\mu}_{p}^{(1)}}{r_1^{7-p}}\right)\cr
    =&-\frac{\tilde{\mu}_{p}^{(1)}(7-p)}{r_1^{8-p}}\tilde{\star}_{9-p}^{(1)} d r_1\cr 
    =&-\frac{\tilde{\mu}_{p}^{(1)}(7-p)}{r_1^{9-p}} \tilde{\star}_{9-p}^{(1)} \lp x^{p+1} dx^{p+1} + x^{p+2}dx^{p+2}+ ... + x^9dx^9 \rp\cr   
    =&\frac{\tilde{\mu}_{p}^{(1)}(7-p)}{r_1^{9-p}} \bigg(x^{p+1} dx^{p+2} \wedge dx^{p+3}\wedge ... \wedge dx^9\cr
    &\qquad \qquad- x^{p+2}dx^{p+1} \wedge dx^{p+3} \wedge ... \wedge dx^9 \cr
    &\qquad \qquad+ \,...\cr
    &\qquad \qquad+ (-1)^p \left(x^9 dx^{p+1} \wedge dx^{p+2} \wedge ...\wedge dx^8\right)\bigg)\,.
\end{align}
Explicitly we find the following component that we will use below
\begin{equation}
    \left(F_{8-p}^{(1)}\right)_{(p+1)(p+3)(p+4) \ldots 9} = -\frac{\tilde{\mu}_{p}^{(1)}(7-p)}{r_1^{9-p}}\,  x^{p+2}\,.
\end{equation}
$F_2^{(2)}$ can be obtained by exchanging $x^{p+1},x^{p+2}$ with $x^{p-1},x^p$ in equation \eqref{eq:F21expl}. In particular, it has the component
\begin{equation}
    \left(F_{8-p}^{(2)}\right)_{(p-1)(p+3)(p+4) \ldots 9} = -\frac{\tilde{\mu}_{p}^{(2)}(7-p)}{r_2^{9-p}}\,  x^{p}\,.
\end{equation}
Note that the above $F_{8-p} = F_{8-p}^{(1)} + F_{8-p}^{(2)} + F_{8-p}^{(c)}$ is the most generic and exact solution to the Bianchi identity in equation \eqref{eq:Bianchi}. It is independent of our particular metric Ansatz since the warp factors do not appear.

\subsection{The Einstein and dilaton equations}

Now we can look at Einstein's equations from equation \eqref{eq:GR} that reduce to 
 \begin{align}\label{eq:Einstein}
    R_{ab} = &-2\nabla_a\partial_b\phi + \frac{1}{4}g_{ab}(2g^{cd}\partial_c\phi\partial_d\phi-\nabla^2\phi) \\
    &e^{2\phi} \left( \frac{1}{2(1+\delta_{(8-p))5})}|F_{8-p}|^2_{ab} - \frac{7-p}{16(1+\delta_{(8-p)5})}g_{ab}|F_{8-p}|^2\right) + \frac{1}{2}e^\phi(T^{loc}_{ab}-\frac{1}{8}g_{ab}T^{loc})  \nonumber \,.
    \end{align}
Calculating the Ricci scalar for the above metric Ansatz in equation \eqref{eq:metric} we find for $a=p-1$, $b=p+1$ (essentially from equation \eqref{eq:derivatives} but also via an explicit computation) that
\begin{equation}
    R_{(p-1)(p+1)} = x^{p-1} x^{p+1} f_R(\rho_1,\rho_2,\rho_T)\,,
\end{equation}
where $f_R(\rho_1,\rho_2,\rho_3)$ is a specific function that one can calculate from the above metric Ansatz in equation \eqref{eq:metric}. The important point is that the entire $R_{(p-1)(p+1)}$ component of the Ricci tensor is proportional to derivatives with respect to $x^{p-1}$ and $x^{p+1}$. This then leads (cf. equation \eqref{eq:derivatives}) to the above prefactor $x^{p-1} x^{p+1}$ in front of $f_R(\rho_1,\rho_2,\rho_3)$.

Likewise we find that the dilaton Ansatz $\phi = \phi(\rho_1,\rho_2,\rho_T)$ leads to
\begin{equation}
    -2\nabla_{p-1}\partial_{p+1}\phi =x^{p-1} x^{p+1} f_\phi(\rho_1,\rho_2,\rho_T)\,.
\end{equation}
Let us assume first that the $F_{8-p}$-flux is simply the superposition of the fluxes from the two single sources as might be expected due to the linearity of the corresponding equation \eqref{eq:2sourcesBianchi}. That means we are setting the closed piece $F_{8-p}^{(c)}$ to zero in the $F_{8-p}$-flux. This also means that neither the other RR-fluxes nor the $H$-flux are sourced.

All non-diagonal entries of the metric in equation \eqref{eq:metric} vanish and the source terms vanish away from the sources as well. Therefore, the off-diagonal entry of the Einstein equation \eqref{eq:Einstein} for $(ab) =(p-1 \, p+1)$ is given by
\begin{align}\label{eq:Einsteinoffdiagonal}
    R_{p-1 \, p+1} &= -2\nabla_{p-1}\partial_{p+1}\phi + \frac12 e^{2\phi}|F_{8-p}|^2_{p-1\,p+1}\cr
    x^{p-1} x^{p+1} f_R(\rho_1,\rho_2,\rho_T) &= x^{p-1} x^{p+1} f_\phi(\rho_1,\rho_2,\rho_T)\cr
    &\quad + e^{2\phi} \frac{1}{2(p-1)!} F_{8-p,p-1\,a_1...a_{7-p}} g^{a_1b_1}...\,g^{a_{7-p}b_{7-p}} F_{8-p,p+1\,b_1...b_{7-p}}\cr
    &= x^{p-1} x^{p+1} f_\phi(\rho_1,\rho_2,\rho_T)\cr
    &\quad+ e^{2\phi} \frac{1}{2} F_{8-p,p-1\,{p+3}...{9}} g^{p+3 p+3}...\, g^{99}F_{8-p, p+1 \, {p+3}...{9}}\cr
    &= x^{p-1} x^{p+1} f_\phi(\rho_1,\rho_2,\rho_T)\cr
    &\quad+ e^{2\phi} \frac{1}{2} \tilde{\mu}_{p}^{(2)} (7-p) \frac{x^{p}}{{r_2}^{9-p}} e^{-2(7-p)A_4}\tilde{\mu}_{p}^{(1)}(7-p) \frac{x^{p+2}}{{r_1}^{9-p}}\,.
\end{align}
We rewrite this as
\be
x^{p-1} x^{p+1} (f_R - f_\phi) = x^{p} x^{p+2} \lp \frac{ e^{2\phi-2(7-p)A_4}}{2} \frac{\tilde{\mu}_{p}^{(1)} (7-p)}{{r_1}^{9-p}} \frac{\tilde{\mu}_{p}^{(2)}(7-p)}{{r_2}^{9-p}}\rp\,.
\ee
The above equation has to be true for all $x^{p-1}, x^p, x^{p+1}, x^{p+2}$. In particular, the left-hand-side is odd under the sign flips $x^{p-1} \to - x^{p-1}$ or $x^{p+1} \to - x^{p+1}$ and even under the sign flips $x^{p} \to - x^{p}$ or $x^{p+2} \to - x^{p+2}$. Since $\phi$, $A_4$, $r_1$ and $r_2$ are all functions of $(\rho_1,\rho_2,\rho_T)$, we find that the symmetry properties of the right-hand-side are exactly opposite. This means the left- and right-hand-side have to vanish independently. Since the dilaton and the component $e^{2A_4}$ of the diagonal metric cannot vanish everywhere we conclude that the vanishing of the right-hand-side implies that 
\begin{equation}\label{eq:contradiction}
    \tilde{\mu}_{p}^{(1)} \tilde{\mu}_{p}^{(2)} = 0\,.
\end{equation}
The above equation implies that one of the two sources is absent. Or, if we insist that both of the intersecting sources are present, we have shown that there is no solution to the supergravity equations of motion for our two intersecting localized sources with our generic diagonal metric Ansatz. To make this proof fully general, we have to allow for the closed piece $F_{8-p}^{(c)}$ as solution to the Bianchi identity \eqref{eq:2sourcesBianchi}. Then this closed form piece can source the $H$-flux and other RR-fluxes via the equations of motion for the fluxes given above in \eqref{eq:EM}. Thus, in order to give a full proof we have to actually allow for all possible RR-fluxes and the most generic $H$-flux compatible with our $SO(p-2,1)\times SO(2) \times SO(2) \times SO(7-p)$ symmetry group. This makes the Einstein and dilaton equations too complicated to analyze directly. Therefore, in the next subsection we study the spinor equations and show that there is indeed no supersymmetric localized solution to the supergravity equations of motion.

\subsection{Spinor equations for the most generic fluxes}

Let us discuss the most generic forms that are invariant under the assumed symmetry group $SO(p-2,1)\times SO(2) \times SO(2) \times SO(7-p)$ for the backreacted solution. Since the first factor $SO(p-2,1)$ is enhanced to the full Poincar\'e group, the only invariant forms are the always present 0-form and its Hodge dual which is the volume form that is proportional to $dx^0 \w dx^1 \w \ldots \w dx^{p-2}$. The other three spaces all have an $SO(n)$ symmetry so we can discuss them together: In addition to the 0-form and the dual volume form, there are two more forms. There is one 1-form which is $d$ acting on the radial coordinates, $d\rho_1$, $d\rho_2$, $d\rho_T$ in our case, and then there is the dual $(n-1)$-form. For an $SO(2)$ symmetry this would be another 1-form, which we denote $d\theta_1$ and $d\theta_2$, where $(\rho_i,\theta_i)$ are simply polar coordinates. For the $SO(7-p)$-symmetry we would go to spherical coordinates $(\rho_T, \theta_T^{(1)},\theta_T^{(2)}, \ldots,\theta_T^{(6-p)})$ and an invariant $(6-p)$-form is given by $\sin(\theta_T^{(1)})^{5-p} \sin(\theta_T^{(2)})^{4-p}\ldots \sin(\theta_T^{(4-p)})^{2} \sin(\theta_T^{(5-p)}) \, d \theta_T^{(1)} \wedge \theta_T^{(2)} \wedge \ldots \wedge \theta_T^{(6-p)}$. Lastly, we note that all functions like the metric, the warp factors, the dilaton or the prefactors that appear in front of the forms when spelling out the fluxes, can only depend on $\rho_1, \rho_2, \rho_T$ due to the preserved symmetry.

Let us give a concrete example to clarify the above discussion. We choose $p=6$ and want to find a localized solution that describes two O6-planes (or D6-branes) that extend along the directions $(x^0,x^1,x^2,x^3,x^4,x^5,x^6)$ and $(x^0,x^1,x^2,x^3,x^4,x^7,x^8)$, respectively. We take them to be localized at the origin in their transverse spaces. We assume that the metric is given by equation \eqref{eq:metric} above for $p=6$. We take all warp factors and the dilaton to be functions of the three variables 
\be
\rho_1=\sqrt{(x^5)^2+(x^6)^2}\,,\qquad \rho_2=\sqrt{(x^7)^2+(x^8)^2}\,, \qquad \rho_T =x^9\,.
\ee
We then make the most generic flux Ansatz that is compatible with an $SO(4,1)\times SO(2) \times SO(2)$ symmetry group
\ba
F_{2} &=& F_{2}^{(1)} + F_{2}^{(2)} + F_{2}^{(c)}\,,\cr
F_{2}^{(1)} &=& \tilde{\star}_{3}^{(1)} d \left(\frac{\tilde{\mu}_{6}^{(1)}}{\sqrt{\rho_1^2 + \rho_T^2}}\right)\,,\cr
F_{2}^{(2)} &=& \tilde{\star}_{3}^{(2)} d \left(\frac{\tilde{\mu}_{6}^{(2)}}{\sqrt{\rho_2^2 + \rho_T^2}}\right)\,,\cr
F_{2}^{(c)} &=& \sum_{i=1}^{10} f_2^{(i)} (\rho_1, \rho_2, \rho_T)\, Y^2_i\,,\cr
F_{4} &=& \sum_{i=1}^{5} f_4^{(i)}(\rho_1, \rho_2, \rho_T)\,  Y^4_i\,,\cr
H &=& \sum_{i=1}^{10} h^{(i)} (\rho_1, \rho_2, \rho_T)\, Y^3_i\,.
\ea
Here the $f_2^{(i)}$, $f_4^{(i)}$, $h^{(i)}$ are unknown functions and the $Y^2_i$, $Y^3_i$, $Y^4_i$ denote the invariant and closed forms that form a basis of invariant forms. Since the $f_2^{(i)}(\rho_1, \rho_2, \rho_T)$ are generic functions and the $Y^2_i$ include for example $d\theta_1 \w d\theta_2$, this Ansatz does not yet satisfy $dF_{2}^{(c)}=0$. We furthermore allow for a constant and non-zero $F_0$. The two 16 component spinors that are present in 10d flat space are constrained due to the presence of the O6-planes (or D6-branes) and have to satisfy
\ba
\epsilon_2 = \Gamma_{0123456} \epsilon_1\,,\cr
\epsilon_2 = \Gamma_{0123478} \epsilon_1\,.
\ea
This breaks one quarter of the supersymmetry and leaves us with 8 real independent spinor components. The fully backreacted solution should preserve these eight supercharges. We therefore assume that these eight spinors are independent (and also functions of $(\rho_1,\rho_2,\rho_T)$). 

We now demand that there is a supersymmetric solution and therefore demand that the spinor transformations in equation \eqref{eq:spinortrafos} satisfy $\delta_\epsilon \psi_a = \delta_\epsilon \lambda = 0$. This leads directly to $F_0 = F_4 = H = 0$, while $F_2$ has to be of the following form
\ba\label{eq:F2solO6}
F_2 &=& \tilde{\star}_{3}^{(1)} d \left(\frac{\tilde{\mu}_{6}^{(1)}}{\sqrt{\rho_1^2 + \rho_T^2}}\right) + \tilde{\star}_{3}^{(2)} d \left(\frac{\tilde{\mu}_{6}^{(2)}}{\sqrt{\rho_2^2 + \rho_T^2}}\right) \nn \\[5pt]
&&+f_2^{(3)}(\rho_1,\rho_2,\rho_T) d\theta_1\wedge d\rho_1 + f_2^{(8)}(\rho_1,\rho_2,\rho_T) d\theta_1\wedge d\rho_T \nn \\[5pt]
&&+ f_2^{(7)}(\rho_1,\rho_2,\rho_T) d\theta_2\wedge d\rho_2 + f_2^{(9)}(\rho_1,\rho_2,\rho_T) d\theta_2\wedge d\rho_T \nn \\[5pt]
&=& \lp f_2^{(3)}(\rho_1,\rho_2,\rho_T) + \frac{\mu_{6,1} \rho_1 \rho_T}{(\rho_1^2+\rho_T^2)^{\frac32}} \rp d\theta_1\wedge d\rho_1 \nn \\[5pt]
&&+\lp f_2^{(8)}(\rho_1,\rho_2,\rho_T) - \frac{\mu_{6,1} \rho_1^2}{(\rho_2^2+\rho_T^2)^{\frac32}} \rp d\theta_1\wedge d\rho_T \nn \\[5pt]
&&+ \lp f_2^{(7)}(\rho_1,\rho_2,\rho_T) + \frac{\mu_{6,2} \rho_2 \rho_T}{(\rho_1^2+\rho_T^2)^{\frac32}}\rp d\theta_2\wedge d\rho_2 \nn \\[5pt]
&& +\lp f_2^{(9)}(\rho_1,\rho_2,\rho_T) - \frac{\mu_{6,2} \rho_2^2}{(\rho_2^2+\rho_T^2)^{\frac32}} \rp d\theta_2\wedge d\rho_T\,.
\ea
Recall that we have made a fully generic Ansatz for the closed piece in $F_2$ and we have not yet imposed that it is actually closed. 

Let us briefly discuss the above solution in equation \eqref{eq:F2solO6}. We see that without imposing the Bianchi identities and equations of motions for the fluxes we can only have a very limited number of flux components $f_2^{(i)}(\rho_1,\rho_2,\rho_T)$ in addition to the source terms. These extra flux components actually combine with the source terms which makes perfect sense. For example, we know that there are solutions for a single source and we can for example use the $f_2^{(i)}(\rho_1,\rho_2,\rho_T)$ to remove one of the sources and then we actually reproduce the result for a single source discussed above in subsection \ref{ssec:singlesource}. Here however, we are interested in solutions that describe two intersecting sources and we therefore do not want to cancel any source terms. We therefore proceed to study the remaining equations of motion.

We want that the source terms containing $\mu_{6,1}$ and $\mu_{6,2}$ give rise to the delta function sources and that the rest is closed (see the discussion around equation \eqref{eq:2sourcesBianchi} above). Thus, we have to demand that $dF_2=0$ away from the source and therefore we find that 
\ba\label{eq:dF2zero}
\partial_{\rho_2} f_2^{(3)}(\rho_1,\rho_2,\rho_T) &=& 0\,,\cr
\partial_{\rho_1} f_2^{(7)}(\rho_1,\rho_2,\rho_T) &=& 0\,,\cr
\partial_{\rho_2} f_2^{(8)}(\rho_1,\rho_2,\rho_T) &=& 0\,,\cr
\partial_{\rho_1} f_2^{(9)}(\rho_1,\rho_2,\rho_T) &=& 0\,.
\ea
Additionally, the spinor equations in \eqref{eq:spinortrafos} did not only set most of the flux components to zero but they also fixed the first derivatives of the warp factors via the spin connection term in  $\delta_\epsilon \psi_a=0$ and the first derivatives of the dilaton via $\delta_\epsilon \lambda=0$. Concretely, they fix $\partial_{\rho_1} e^{A_2(\rho_1,\rho_2,\rho_T)}$ and $\partial_{\rho_2} e^{A_2(\rho_1,\rho_2,\rho_T)}$ to be two different functions of the warp factors, the dilaton, the $f_2^{(i)}(\rho_1,\rho_2,\rho_T)$ and the source terms
\ba
\partial_{\rho_1} e^{A_2(\rho_1,\rho_2,\rho_T)} = F_1(e^{A_i}, e^\phi, f_2^{(i)},\rho_1,\rho_2,\rho_T)\,,\cr
\partial_{\rho_2} e^{A_2(\rho_1,\rho_2,\rho_T)} = F_2(e^{A_i}, e^\phi, f_2^{(i)},\rho_1,\rho_2,\rho_T)\,.
\ea
Now we can impose the conditions above in equation \eqref{eq:dF2zero} and the following consistency condition
\ba
0&=&\partial_{\rho_2} \partial_{\rho_1} e^{A_2(\rho_1,\rho_2,\rho_T)} -\partial_{\rho_1} \partial_{\rho_2} e^{A_2(\rho_1,\rho_2,\rho_T)}\cr
&=&\partial_{\rho_2} F_1(e^{A_i}, e^\phi, f_2^{(i)},\rho_1,\rho_2,\rho_T)-\partial_{\rho_1} F_2(e^{A_i}, e^\phi, f_2^{(i)},\rho_1,\rho_2,\rho_T)\cr
&=& \frac{e^{A_2-2A_4+2\phi}}{2 \rho_1 \rho_2} \left( f_2^{(8)}(\rho_1,\rho_T)-\frac{\mu_{6,1}\rho_1^2}{\left(\rho_1^2 +\rho_T^2\right)^{\frac32}}\right) \left(f_2^{(9)}(\rho_2,\rho_T)-\frac{\mu_{6,2} \rho_2^2}{\left(\rho_2^2+\rho_T^2\right)^{\frac32}}\right)\,.\quad
\ea
Since the prefactor in the above equation cannot vanish everywhere, we see that
\be\label{eq:contradiction2}
\left( f_2^{(8)}(\rho_1,\rho_T)-\frac{\mu_{6,1}\rho_1^2}{\left(\rho_1^2 +\rho_T^2\right)^{\frac32}}\right) \left(f_2^{(9)}(\rho_2,\rho_T)-\frac{\mu_{6,2} \rho_2^2}{\left(\rho_2^2+\rho_T^2\right)^{\frac32}}\right) = 0\,.
\ee
This shows that there is no fully localized solution with our generic diagonal metric Ansatz. The above equation requires us to at least partially remove (or smear) one of the sources.

Let us pursue the above further by setting without loss of generality
\be
f_2^{(9)}(\rho_2,\rho_T)=\frac{\mu_{6,2} \rho_2^2}{\left(\rho_2^2+\rho_T^2\right)^{\frac32}}\,.
\ee
This cancels the last term in $F_2$ above in equation \eqref{eq:F2solO6} and the closure $dF_2=0$ then imposes the additional constraint that
\be
f_2^{(7)}(\rho_1,\rho_2,\rho_T) = \frac{\mu_{6,2} \rho_2^2}{(\rho_2^2+\rho_T^2)^{\frac32}} + f(\rho_2)\,,
\ee
where $f(\rho_2)$ is an undetermined function. With that $F_2$ becomes
\ba
F_2 &=& \lp f_2^{(3)}(\rho_1,\rho_T) + \frac{\mu_{6,1} \rho_1 \rho_T}{(\rho_1^2+\rho_T^2)^{\frac32}} \rp d\theta_1\wedge d\rho_1 \cr
&&+\lp f_2^{(8)}(\rho_1,\rho_T) - \frac{\mu_{6,1} \rho_1^2}{(\rho_2^2+\rho_T^2)^{\frac32}} \rp d\theta_1\wedge d\rho_T \cr
&& +f(\rho_2) d\theta_2\wedge d\rho_T\,.
\ea
So, we have effectively removed the second source completely. Actually the equations of motion for $F_2$ fix $f(\rho_2) = c \rho_2$ and using that in the solution to the spinor equations, we find that all derivatives of the warp factors and the dilaton with respect to $\rho_2$ vanish: $\partial_{\rho_2} e^{A_i} = \partial_{\rho_2} e^\phi=0$. This is indicative of a smeared source and we indeed see from 
\be
dF_2 \supset d(f(\rho_2) d\theta_2\wedge d\rho_T) = c\, d\rho_2 \wedge d\theta_2\wedge d\rho_T\,,
\ee
that we can have at best a smeared second source in which the delta function source (see equation \eqref{eq:2sourcesBianchi}) is replaced with the constant $c$. Thus, in addition to proving the absence of a solution with two fully localized sources, our equations pass consistency checks and do not forbid  solutions with partially smeared sources.\\

We have repeated the above analysis for two intersecting sources with $p=1,2,3,4,5$ and explicitly reproduced the same absence of localized solutions. This might have been expected from T-duality invariance of type II string theory, however, there is an important subtlety: If we have an O$p$-plane (or a D$p$-brane) in flat space, then we can T-dualize along any of its worldvolume directions. The reason is that the dilaton, metric and everything else does not depend on these coordinates. This leads to an O$(p-1)$-plane (or a D$(p-1)$-brane) that is actually smeared over the direction we T-dualized. Similarly, we cannot T-dualize along a transverse direction since these are not isometries. We would have to first smear the source along this transverse direction and then we can T-dualize to get a $(p+1)$-dimensional source. So, strictly speaking we cannot use T-duality invariance in the strict sense and therefore we checked the equations for each $p=1,2,3,4,5,6$ explicitly.

\section{Discussion and Outlook}\label{sec:Outlook}
The above surprising result, that shows the absence of localized supergravity solutions for intersecting objects in flat space, raises many important questions: Does the result also hold for a generic non-diagonal metric? Can such setups be described explicitly in the full string theory? Does our result carry over to compactifications? In this section we will briefly discuss these questions. However, we will not be able to answer them and leave many avenues for further research.

First it seems clear that intersecting sources can arise in string theory and corresponding solutions will exist. Our two intersecting O$p$-planes can arise from a single orientifold projection combined with a $\mathbb{Z}_2$ orbifold of flat space. For example, we can do an orientifold involution consisting of the worldsheet parity operator $\Omega_p$ and a spatial involution that flips the signs of $x^7,x^8,x^9$. This leads to a single O6-plane localized at $x^7=x^8=x^9=0$. Doing a $\mathbb{Z}_2$ orbifold that flips the signs of $x^5, x^6, x^7, x^8$ then introduces a second O6-plane localized at $x^5=x^6=x^9=0$. In principle one should be able to study the full string theory on such an orientifolded orbifold of flat space. Supergravity as a low energy approximation of the full string theory might simply not allow for a solution because we neglect higher derivative corrections, string loop corrections and/or did not include the full spectrum of the string states. 

For the case of intersecting stacks of D-branes in particular we neglected all the open strings on the D-branes. These open strings give rise to gauge theories and one can study the dynamics of these gauge theories. It is possible that the gauge dynamics leads to a (partial) smearing of the D-branes and partially smeared solutions do certainly exist, see for example \cite{Youm:1999ti, Marolf:1999uq, Gomberoff:1999ps, Arapoglu:2003ah}. Some of these papers also discuss the near core (near horizon) limit of these brane setups and manage to find localized solutions in this limit. For the particular case of two intersecting D6-branes or O6-planes one can also try to lift things to M-theory and try to find a solution in 11d supergravity. Such a lift of two intersecting D6-branes was discussed in \cite{Uranga:2002ag}

Let us mention that it is known that multiplying together the two harmonic functions (warp-factors) for the two sources cannot solve the localized equations of motion but rather requires smearing (see for example \cite[eqns. (1)-(2)]{Youm:1999ti} and references therein). We reproduce the same result with a generic diagonal metric Ansatz. A loophole to our findings is exploited by the only (to us) known fully localized supergravity solution of two intersecting branes \cite{Gauntlett:1996pb}. The two intersecting NS5 branes in this setup have no mutually transverse direction since they extend along 012345 and 016789. In our equations we crucially use the fact that there are $7-p>0$ transverse directions.\footnote{Recall that here we restrict ourselves to $1 \leq p \leq 6$. It thus might be possible to write down fully localized solutions for $p=7$ but such setups are better described in F-theory \cite{Vafa:1996xn}.} It would be interesting to study further brane setups without mutually transverse directions.

We crucially assumed here that there is an unbroken $SO(p-2,1)\times SO(2) \times SO(2) \times SO(7-p)$ symmetry group. This seems to be justified for static objects or in the probe limit but it is possible that the dynamics of the D-branes (or the dynamics of O-planes at strong coupling) could break this symmetry group. It would therefore be interesting to see whether one can relax the requirement of this large unbroken symmetry group. Here let us note that \cite{Patino:2004cz} discusses the D3/D5-brane intersection, where the D5 brane extends along $x^0,x^1,x^2,x^3,x^4,x^5$ and the D3-brane along $x^0,x^1,x^2,x^6$. In this case the authors only assume the presence of an $SO(2,1)$ Poincar\'e group and an $SO(3)$-symmetry group in the mutually transverse $x^7,x^8,x^9$ directions. While solving the equations of motion they discover the necessary presence of an extra $SO(3)$-symmetry acting on the $x^3,x^4,x^5$ directions, before they find that no localized solution exists. Thus, it is conceivable that our result might still hold even if we were to give up the $SO(2)\times SO(2)$ symmetry and/or allow for a non-diagonal metric along the corresponding directions. It would be interesting to check this explicitly in particular given that the orbifold blow ups discussed recently in  \cite{Junghans:2023yue} would break this $SO(2)\times SO(2)$ symmetry. In our setup one could glue in a $\mathbb{P}^1$ to remove the orbifold singularity and this would correspond to giving a non-zero vev to the K\"ahler modulus that controls its size. However, it is unclear to us that this would happen dynamically and what could fix the scale of a non-zero vev for the K\"ahler modulus.

It is a far stretch to go from our setup of two intersecting sources in flat space to a full compactification of 10d supergravity like the massive type IIA flux compactifications discussed in \cite{DeWolfe:2005uu, Camara:2005dc}. However, we note here that the two papers \cite{Junghans:2020acz, Marchesano:2020qvg} only worked to first order in the sources, i.e., in our language to the first order in the $\tilde{\mu}_p$. This means that contradictions like equations \eqref{eq:contradiction} or \eqref{eq:contradiction2} above that are quadratic would not be visible when working at linear order. It would be therefore of great importance to extend the work of \cite{Junghans:2020acz, Marchesano:2020qvg} to higher order. Already in the simplest case of a toroidal compactification we note that the preserved symmetry group gets dramatically reduced and it is conceivable that then localized solutions exist. We plan to study this as well as a generic off-diagonal metric Ansatz in the future.

\section*{Acknowledgments}
We like to thank David Andriot, Mariana, Gra\~na, Daniel Junghans, Douglas Smith, Vincent Van Hemelryck and Thomas Van Riet for useful discussions. We are indebted to Daniel Junghans for invaluable feedback on an earlier version of this draft. This work is supported in part by the NSF grant PHY-2210271. T.W. would like to thank the LAPTh - CNRS for hospitality during part of this work. His visit was made possible thanks to the AAP USMB dSCordes (2023). This research was supported in part by grant NSF PHY-2309135 to the Kavli Institute for Theoretical Physics (KITP).

\bibliographystyle{JHEP}
\bibliography{refs}

\end{document}